%% file: xnsv.tex
\documentclass[aps,prl,twocolumn,showpacs,10pt,superscriptaddress,groupedaddress, footinbib]{revtex4-1}

\usepackage{amsmath,bm,amssymb}
\usepackage{amsbsy}
\usepackage{autobreak}
\usepackage{endnotes}
\let\footnote=\endnote

\usepackage{graphicx}
\usepackage{psfrag}
\usepackage[caption=false]{subfig}
\usepackage{color}
\definecolor{urlblue}{rgb}{0.2,0.4,0.7}
 \definecolor{citegreen}{rgb}{0,0.6,0.2}
\usepackage[caption=false]{subfig}
\usepackage{color}
\definecolor{urlblue}{rgb}{0.2,0.4,0.7}
\definecolor{citegreen}{rgb}{0,0.6,0.2}
\definecolor{linkred}{rgb}{0.9,0.2,0.1}
\usepackage{hyperref}
\hypersetup{
colorlinks=true, citecolor=citegreen, linkcolor=blue,urlcolor=urlblue}
\usepackage[table]{xcolor}
\usepackage{tabularx}
\newcolumntype{P}[1]{>{\centering\arraybackslash}p{#1}}
\usepackage{multirow, caption, booktabs}
\usepackage{slashed}
\usepackage{array}
\newcolumntype{P}[1]{>{\centering\arraybackslash}p{#1}}
\allowdisplaybreaks

\def\zo{\overline{z}_1}
\def\zt{\overline{z}_2}

\definecolor{linkred}{rgb}{0.9,0.2,0.1}
\usepackage{hyperref}
\hypersetup{
colorlinks=true, citecolor=citegreen, linkcolor=blue,urlcolor=urlblue}

\usepackage[table]{xcolor}
\usepackage{tabularx}
\newcolumntype{P}[1]{>{\centering\arraybackslash}p{#1}}
\usepackage{multirow, caption, booktabs}

\usepackage{slashed}
\usepackage{array}
\newcolumntype{P}[1]{>{\centering\arraybackslash}p{#1}}

\allowdisplaybreaks

\def\zo{\overline{z}_1}
\def\zt{\overline{z}_2}

\begin{document}


\title{Next-to-soft corrections for Drell-Yan and Higgs boson rapidity distributions beyond N$^3$LO
}

\author{A.H. Ajjath$^{\scriptstyle 1}$}
\email{ajjathah@imsc.res.in}
\author{Pooja Mukherjee,$^{\scriptstyle 1}$}
\email{poojamukherjee@imsc.res.in}
\author{V. Ravindran,$^{\scriptstyle 1}$}
\email{ravindra@imsc.res.in}
\author{Aparna Sankar,$^{\scriptstyle 1}$}
\email{aparnas@imsc.res.in}
\author{Surabhi Tiwari$^{\scriptstyle 1}$}
\email{surabhit@imsc.res.in}
\affiliation{$^{\scriptstyle 1}$The Institute of Mathematical Sciences, HBNI, Taramani,
 Chennai 600113, India}

\date{\today}

\begin{abstract}
We present a formalism that resums both soft-virtual (SV) and next to SV (NSV) contributions to all orders in perturbative QCD for the rapidity distribution of any colorless particle produced in hadron colliders. Using the state-of-the-art results, we determine the complete NSV contributions to third order in strong coupling constant for the rapidity distributions for Drell-Yan and also for Higgs boson in gluon fusion as well as bottom quark annihilation. Using our all order $z$ space result, we show how the NSV contributions can
be resummed in two-dimensional Mellin space.
\end{abstract}
\maketitle
\textit {Introduction}.---Accurate measurement of observables at the Large Hadron Collider (LHC) and 
their precise theoretical predictions,    
provide an opportunity to test the Standard Model (SM) with unprecedented
accuracy thereby constraining beyond the SM (BSM) scenarios.  
One of the cleanest observables at the LHC  is Drell-Yan (DY) production \cite{Drell:1970wh} of 
on-shell vector bosons Z and $W^\pm$ or a pair of leptons and hence
it has received enormous attention from the theory community.
Measurements \cite{Affolder:2000rx,Abe:1998rv,CMS:2014jea} of inclusive and differential rates of DY production are used as a standard candle to calibrate the detectors and also
to fit the non perturbative parton distribution functions (PDF) \cite{Gao:2013xoa,Harland-Lang:2014zoa,Ball:2014uwa,
Butterworth:2015oua,Alekhin:2017kpj}.
Any  deviation from the SM predictions can provide
crucial information  to BSM scenerios, such as R-parity violating supersymmetric models,
models with $Z'$ and large extra-dimension models \cite{ArkaniHamed:1998rs,Randall:1999ee}.
Similarly, the ongoing measurements of inclusive and differential cross sections 
\cite{Aad:2012tfa,Chatrchyan:2012xdj}, along with  
the theoretical predictions \cite{Anastasiou:2015vya} on
strong and electroweak radiative corrections help us to probe the symmetry-breaking  mechanism 
and the coupling of the Higgs boson with other SM partcles.  
This is possible owing to the third order QCD  predictions for 
DY production 
\cite{Duhr:2020seh,Ahmed:2014cla}
and Higgs boson productions 
in gluon fusion \cite{Anastasiou:2015vya,Li:2014bfa,Catani:2014uta}
and bottom quark annihilation
\cite{Duhr:2019kwi,Ahmed:2014era}.\\
Like inclusive rates, differential ones also get large contributions from logarithms from phase space 
boundaries of the final state particles, thus spoiling the reliability of the fixed-order predictions.  
These large logarithms can be summed up to all orders in perturbation theory. In the seminal works  of Sterman \cite{Sterman:1986aj} and of Catani and Trentadue \cite{Catani:1989ne}, resummation of leading large logs for the inclusive rates in Mellin space and also to  differential $x_F$ distribution \cite{Catani:1989ne} using double Mellin moments were achieved.   
Using factorization properties of differential cross sections and renormalisation group (RG) invariance, an all order $z$-space formalism was also developed in \cite{Ravindran:2006bu}, to study  the threshold-enhanced contribution to rapidity distribution of any colorless particle. The formalism was also applied to $Z$ and $W^{\pm}$ \cite{Ravindran:2007sv} and also to DY and Higgs production at N$^3$LO level \cite{Ahmed:2014uya,Ahmed:2014era}. In \cite{Banerjee:2017cfc}, the same formalism \cite{Ravindran:2006bu} was used to study threshold resummation of rapidity distribution of Higgs bosons and later  to DY production \cite{Banerjee:2018vvb}.
For different approaches and their applications, see \cite{
Laenen:1992ey,Sterman:2000pt,Mukherjee:2006uu,Bolzoni:2006ky,Becher:2006nr,Becher:2007ty,Bonvini:2014qga,Ebert:2017uel,Cacciari:2001cw}.\\
Besides the  threshold logarithms, contributions from subleading logarithms 
are also present in all the partonic channels beyond leading order in perturbation theory. 
These subleading logarithms demonstrate perturbative behaviour similar to those from threshold region, which allows one to study their all order structure.
Such logarithms do appear in inclusive reactions and there have been remarkable progress to understand them.  
See, \cite{Laenen:2008ux,Laenen:2010uz,Bonocore:2014wua,Bonocore:2015esa,Bonocore:2016awd,DelDuca:2017twk,Bahjat-Abbas:2019fqa,Soar:2009yh,Moch:2009hr,deFlorian:2014vta,Beneke:2018gvs,Bahjat-Abbas:2019fqa,Beneke:2019mua,Beneke:2019oqx} for more details. Recently, in a series of articles \cite{Ajjath:2020ulr,Ajjath:2020sjk}, 
we studied variety of inclusive reactions to understand these subleading logarithms and found a systematic way to sum them up
 to all orders in $z$ as well as in  Mellin $N$ spaces.  The latter
provides resummed prediction in $N$ space for subleading logarithms similar to that of threshold ones.  

{\color{black}The differential distributions often show richer logarithmic structure due to
multi-dimensional space (spanned by $z_l$ or $N_l$) making it a challenging task to understand the all order structure.}
In the present letter,  using factorisation properties of physical observables and RG invariance,
we have achieved the task of organising the subleading logarithms in a systematic
fashion that is suitable for summing them up to all orders in perturbation theory both in $z_l$ and $N_l$ spaces.

\textit {Theoretical framework}.--- In QCD improved parton model, the rapidity distribution of a colorless state $F$ 
in hadron-hadron collision is given by 
\begin{eqnarray}\label{sighad}
{d \sigma^c\over dy } &=&
\sigma^c_{\rm B}(\tau, q^2) 
\sum_{a,b=q,\overline q,g}
\int_{x_1^0}^1 {dz_1 \over z_1}\int_{x_2^0}^1 {dz_2 \over z_2}~ 
f_{a}\left({x_1^0 \over z_1},\mu_F^2\right)
\nonumber \\ 
&& \times f_{b}\left({x_2^0\over z_2}, \mu_F^2\right)
\Delta^c_{d,ab} (z_1,z_2,q^2,\mu_F^2,\mu_R^2)\,,
\end{eqnarray}
where $\sigma^c_B(\mu_R^2)=\sigma^c_B(x_1^0,x_2^0,q^2,\mu_R^2)$ is the born cross section and $\mu_R$ is the ultraviolet (UV) renormalization scale. The scaling variables $x_l^0~(l=1,2)$ are defined through hadronic rapidity $y$:~ $y={1 \over2} \ln(p_2.q/p_1.q)={1 \over 2 } \ln\left({x_1^0/x_2^0}\right)$ and $\tau=q^2/S=x_1^0 x_2^0$.  Here $q$ denotes the momentum of colorless state F and  $S=(p_1+p_2)^2$ is the hadronic center of mass energy, with $p_l~(l=1,2)$ the momenta of incoming hadrons.
 For $F$ being state of a pair of leptons $\sigma^c = d\sigma^{q}(\tau,q^2,y)/dq^2$, i.e, its invariant mass distribution, whereas for the Higgs production in gluon fusion or in bottom quark annihilation 
 $\sigma^c=\sigma^{g,b}(\tau,q^2,y)$ 
respectively. The PDFs $f_c(x_l,\mu_F^2)$ of colliding partons $c=q,\overline q,g,b$ with momentum fractions $x_l~(l=1,2)$
are renormalized at the factorization scale $\mu_F$. 
The partonic coefficient functions (CFs), $\Delta_{d,ab}$,
are perturbatively calculated in QCD in powers of strong coupling constant $a_s(\mu_R^2)=g_s^2(\mu_R^2)/16 \pi^2$ 
and are functions of the scaling variables  $z_l=x_l^0/x_l~(l=1,2)$. They are obtained from the partonic processes through mass factorisation. 
  The UV finite partonic processes contain soft and collinear divergences associated with the soft gluons and collinear partons, beyond leading order in perturbation theory, which can be removed by summing over degenerate final states and by mass factorization. In this letter we restrict ourselves to partonic CFs of only quark-antiquark initiated processes for DY, gluon-gluon and bottom-anti-bottom initiated processes for Higgs productions. We  call them diagonal CFs (dCFs) $\Delta_{d,a\overline a}$~($a=q,g,b$). These dCFs comprise of contributions from $\delta(1-z_l)$ and ${\cal D}_j(z_l)\equiv\big(\frac{\ln^j(1-z_l)}{(1-z_l)}\big)_+$ (namely SV) and the coefficients regular in $z_l$. The leading contributions of the latter near the threshold region $z_l=1$ contain terms of the form ${\cal D}_i(z_l)\ln^k(1-z_j)$ and $\delta(1-z_l) \ln^k(1-z_j)$ with ($l,j=1,2),~(i,k=0,1,\cdots$). We call them next to soft-virtual (NSV) contributions.  In the Mellin $N_l$ space, these terms are of the form of $\ln^k N_j/N_l$ with  $(j,l=1,2),(k=0,1\cdots$). The dominant SV contribution has been studied in the earlier works of one of the authors in \cite{Ravindran:2006bu}. In the following we discuss the NSV contributions of the dCFs in $z_l$ as well as in $N_l$ space. 
 
\textit {Fixed Order Formalism}.--- Using RG invariance and factorization properties of differential dCF \cite{Ravindran:2006bu}, the threshold-enhanced SV and NSV terms of dCF, denoted by
$\Delta^{ \rm SV +\rm NSV }_{d,c}$, is found to exponentiate as
\begin{align}\label{delta}
\Delta^{\rm SV+ \rm NSV}_{d,c} ={\cal C} \exp
\Big({\Psi^c_d(q^2,\mu_R^2,\mu_F^2,\zo,\zt,\epsilon)}\Big)\, \Big|_{\epsilon = 0} \,,
\end{align}
where the function $\Psi^c_d $ is computed in perturbative QCD in  $4+\epsilon$ space-time dimensions and $\zo = 1- z_1$ and $\zt = 1-z_2$ are the shifted scaling variables. It is shown in Eq.(9) of \cite{Ravindran:2006bu} that the UV and IR finite function $\Psi^c_d $ can be decomposed in terms of form factor $F^c$, soft distribution $\Phi^c_d$ and the diagonal Altarelli-Parisi kernels $\Gamma_{cc}$. The soft distribution discussed in \cite{Ravindran:2006bu} using K+G type Sudakov differential equation, accounts for the soft enhancements associated with the real emissions in the production channel and is universal in nature. This universality ensures that $\Phi^c_d$ is only sensitive to the initial legs and is  blind to the hard process under study. In this letter we find that the K+G equation admits solution that can account for next-to-soft contributions as well: 
\begin{align}
\label{eq:Phid}
\Phi^c_d =& \sum_{i=1}^\infty \hat a_s^i \left(q^2 \zo \zt \over \mu^2\right)^{i{\epsilon \over 2}} S_\epsilon^i
\Bigg[ {(i \epsilon)^2 \over 4 \zo \zt } \hat \phi_d^{c,(i)}(\epsilon)
\nonumber \\&
+ {i \epsilon \over 4 \zo } \varphi_{d,c}^{(i)} (\overline z_2,\epsilon)
+ {i \epsilon \over 4 \zt } \varphi_{d,c}^{(i)} (\overline z_1,\epsilon)\Bigg] \,,
\end{align}
where $S_\epsilon = \exp \big(\frac{\epsilon}{2}\big[\gamma_E -\ln(4 \pi) \big]\big)$ with $\gamma_E$ being the Euler-Mascheroni constant.
The first term within the parenthesis accounts for the soft contributions and remaining two terms correspond to next-to-soft contributions. The soft part of the solution was proposed along with the  predictions for Higgs production and DY in \cite{Ravindran:2006bu} till third order, without
$\delta(\overline z_1) \delta(\overline z_2)$ terms. Later on  \cite{Ahmed:2014uya,Ahmed:2014era} gives the complete result for SV.  Through mass factorisation the divergent part of the NSV solution cancels against the collinear singularities from AP kernels and the finite part contributes to dCFs. The coefficients $\varphi_{d,c}^{(i)}$ depend on $\overline z_l$ and $\epsilon$ in such a way that the NSV part is RG invariant provided we sum the series to all orders. 
 In addition, we find that the logarithmic structure of $\Phi_d^c$ and consequently their predictions remain unaltered under the simultaneous transformation of the exponent in first parenthesis and the $z_l$-dependence in $\varphi_{d,c}^{(i)}(z_l,\epsilon)$. The AP kernels satisfy, 
\begin{align}
\mu_F^2 {d \over d \mu_F^2} {\cal{C}} \ln \Gamma_{cc}(\mu_F^2,\overline z_l) = {1 \over 2} P^c(a_s(\mu_F^2),\overline z_l) 
+\delta P^c\,,
\end{align}
where 
\begin{align}
P^c(a_s,\overline z_l) = 2 \bigg({A^c (a_s) \over (\overline z_l)_+} +  B^c (a_s) \delta(\overline z_l) +  L^c(a_s, \overline z_l)\bigg)\,,
\end{align}
with $A^c$ and $B^c$ being the cusp and collinear anomalous dimensions, $L^c(a_s, \overline z_l) \equiv C^c(a_s) \ln(\overline z_l)  + D^c(a_s)$ and the $\delta P^c$ denote NSV and beyond the NSV terms respectively.  We drop $\delta P^c$ throughout. The NSV improved solution $\Phi^c_d$ results in an integral representation of the finite function $\Psi^c_d$ which embeds the all order information of the mass-factorised differential distribution.
\begin{align}
\label{eq:psiint}
\Psi^c_d =& {\delta(\overline z_1) \over 2} \Bigg(\!\!\displaystyle {\int_{\mu_F^2}^{q^2 \overline z_2}
\!\!{d \lambda^2 \over \lambda^2}}\! {\cal P}^c\left(a_s(\lambda^2),\zt\right) 
\!+\! {\cal Q}^c_d\left(a_s(q_2^2),\zt\right)
\!\!\Bigg)_+ 
\nonumber\\&
+ {1 \over 4} \Bigg( {1 \over \overline z_1 }
\Bigg\{{\cal P}^c(a_s(q_{12}^2),\zt ) + {\color{black} 2 }L^c(a_s(q_{12}^2) ,\zt)
\nonumber \\
& + q^2{d \over dq^2} 
\left({\cal Q}^{c}_d(a_s(q_{12}^2 ),\zt) +  {\color{black} 2 }\varphi_{d,c}^f(a_s(q_{12}^2 ),\zt)\right)
\Bigg\}\Bigg)_+
\nonumber\\&
+ {1 \over 2}
\delta(\overline z_1) \delta(\overline z_2)
\ln \Big(g^c_{d,0}(a_s(\mu_F^2))\Big)
+ \overline z_1 \leftrightarrow \overline z_2,
\end{align}
where ${\cal P}^c (a_s, \overline z_l)= P^c(a_s,\overline z_l) - 2 B^c(a_s) \delta(\overline z_l)$, $q_l^2 = q^2~(1-z_l) $ and $q^2_{12}=q^2 \overline z_1 \overline z_2$. The subscript $+$ indicates standard plus distribution. The function ${\cal Q}^c_d$ in \eqref{eq:psiint} is given as
\begin{eqnarray}
{\cal Q}^c_d(a_s,\overline z_l) = {2 \over \overline z_l}  D_d^c(a_s) + 2 \varphi^f_{d,c} (a_s,\overline z_l)\,.
\end{eqnarray}
The splitting function $P^c$ 
and the SV coefficient $ D_d^c$ are known to third order \cite{Banerjee:2017cfc} in QCD.
Here $\varphi_{d,c}^{f}$ constitutes the finite part of  $\varphi_{d,c}^{(i)}$ in \eqref{eq:Phid}
and is parametrized in the following way,
\begin{align}
\label{eq:Phidf}
\varphi_{d,c}^f(a_s(\lambda^2),\overline z_l) &= \sum_{i=1}^\infty \sum_{k=0}^{\infty} \hat  a_s^i \left({\lambda^2 \over \mu^2}\right)^{i\frac{\epsilon}{2}}
S_\epsilon^i 
\varphi^{(i,k)}_{d,c}(\epsilon)\ln^k \overline z_l\,,
\nonumber\\
&= 
\sum_{i=1}^\infty \sum_{k=0}^i a_s^i(\lambda^2) \varphi^{c,(k)}_{d,i} \ln^k \overline z_l\,.
\end{align} 
The upper limit on the sum over $k$ is controlled by the dimensionally regularised Feynman integrals that contribute to order $a_s^i$.   
The constant $g^c_{d,0}$ in \eqref{eq:psiint} results from finite part of the virtual contributions and pure $\delta(\overline z_l)$ terms of $\Phi^c_d$.   
The exponent $\Psi^c_d$ that captures both SV as well as NSV terms to all orders in perturbation theory is one of the
main results of this letter.

\textit {Matching with the Inclusive}.--- The NSV function $\varphi_{d,c}^f$ can be determined at every order in perturbation theory using fixed order
predictions of $\Delta_{d,c}$. Alternatively, 
we can determine $\varphi_{d,c}^f$ from corresponding inclusive cross sections using the relation 
\cite{Ravindran:2006bu}:
\begin{align}
\int_0^1 
dx_1^0 \int_0^1 
dx_2^0 \left(x_1^0 x_2^0\right)^{N-1}
{d \sigma^c \over d y}
=\int_0^1 d\tau~ \tau^{N-1} ~\sigma^c\,,
\label{iden}
\end{align}
where $\sigma^c$ is the inclusive cross section. 
This relation in the large $N$ limit gives
\begin{align}
\label{eq:RapCoeff}
\sum_{i=1}^\infty \hat a_s^i \left({q^2  \over \mu^2} \right)^{i \epsilon \over 2} S_\epsilon^i 
\Big[t_1^{i}(\epsilon) \hat \phi_d^{c,(i)}(\epsilon)
-t_2^{i}(\epsilon) \hat \phi^{c,(i)}(\epsilon)
\nonumber \\ 
+ \sum_{k=0}^{\infty} \Big( t_3^{(i,k)}(\epsilon) \varphi^{(i,k)}_{d,c}(\epsilon)
-t_4^{(i,k)}(\epsilon) \varphi^{(i,k)}_c(\epsilon)\Big)\Big] = 0\,.
\end{align}
Here we keep $\ln^k N$ as well as  ${\cal O}(1/N)$ terms for the determination of the SV and NSV coefficients. The constants $\hat \phi^{c, (i)}$ and $\varphi^{(i,k)}_c$ are the inclusive counterparts to the SV and NSV coefficients respectively which are known to third order in QCD for DY ($c=q$), for Higgs production in gluon fusion ($c=g$) and in bottom quark annihilation ($c=b$) (for NSV see {\cite{Ajjath:2020ulr})}.  
The coefficients are 
\begin{align}
&t_1^i~ = {i\epsilon (2 - i \epsilon) \over 4 N^{i \epsilon}}  \Gamma^2\left(1+i {\epsilon\over 2}\right)\,, \quad 
t_2^i={ i \epsilon  (1 - i \epsilon)  \over 2 N^{i \epsilon}} \Gamma(1+i \epsilon),
\nonumber \\
&t_3^{(i,k)} = \Gamma\left(1+i {\epsilon\over 2}\right){\partial^k \over \partial 
\alpha^k} \left({\Gamma (1+\alpha) \over N^{\alpha + i \epsilon/2}}\right)_{\alpha=i {\epsilon \over 2}},
\nonumber \\
&t_4^{(i,k)} = {\partial^k \over \partial {\hat \alpha }^k} \left({\Gamma (1+\hat \alpha )\over
N^{\hat \alpha}} \right)_{\hat \alpha =i \epsilon}. 
\end{align}

\textit {All order prediction.}---In \cite{Ravindran:2006bu,Ahmed:2014uya,Ahmed:2014era}, we studied the predictive power of SV part of $\Psi_d^c$ to dCFs to all orders using lower order results. 
Here, in particular,  we predict NSV terms of the form $\delta(\overline z_l) \ln^k\overline z_j, n+1 \le k \le 2n-1 $ and ${\cal D}_i(z_l) \ln^k\overline z_j$ for $i,k=0,1,\cdots,n; i+k < 2n-1$ at every order $a_s^n$ provided $\Psi_d^c$ is known to order $a_s^{n-1}$. From $\Psi_d^c,~c=q,b,g$ determined from second order inclusive results \cite{Ajjath:2020ulr}, we obtain for the first time the results for the  third order NSV contributions to dCFs, $\Delta_{d,c}$, for $c=q,b$ and also for $c=g$ \cite{Dulat:2018bfe}. Further, the knowledge of third order results \cite{Ajjath:2020ulr}
for inclusive reactions and using \eqref{eq:RapCoeff} we have determined the NSV coefficients $\varphi_{d,i}^{c,(k)}$ and
dCFs to third order. They will be presented towards the end in concise form. 

\textit {Resummation.}--- Near the hadronic threshold region, $z_l \rightarrow 1$, the PDFs often become large (due to their small momentum fractions)
which allows the threshold contributions from CFs to dominate at every order in $a_s$.  Hence, truncated perturbative predictions
become unreliable.  In Mellin space, these dominant ones show up as order one terms of the form {\color{black}$a_s \beta_0 \ln N_1 N_2$} in the large $N_l$ region at every order. Thanks to all order integral representation for $\Psi_d^c$ in \eqref{eq:psiint} and RG equation of $a_s$, we can resum these terms to all orders.  Defining double Mellin moment of any arbitrary function $F(z_1,z_2)$  by $F_{\vec N}= \int_0^1 dz_1 z_1^{N_1-1}\int_0^1 dz_2 z_2^{N_2-1}F(z_1,z_2)$, we obtain $\Delta_{d,\vec N}^c = \tilde g_{d,0}^c \exp(\Psi_{d,\vec N}^c)$, which can be expanded in terms of $a_s$:  $\Delta_{d,\vec N}^c = \sum_{i=0}^\infty a_s^i(\mu_R^2) \Delta_{d,\vec N}^{c,(i)}$.  
The resummed result for $\Psi_{d,\vec N}^c$ takes the following form:
\begin{align}
\label{eq:PsiN}
\Psi_{d,\vec N}^c = &~~
  \bigg(g_{d,1}^c(\omega)
+\frac{1}{N_1} \overline{g}_{d,1}^c(\omega)\bigg) \ln N_1
\nonumber\\&
+ \sum\limits_{i=0}^\infty a_s^i \bigg( \frac{1}{2}  g_{d,i+2}^c(\omega) + \frac{1}{N_1} \overline{g}_{d,i+2}^c(\omega) \bigg)
\nonumber\\&
 +\frac{1}{N_1}
\sum\limits_{i=0}^{\infty} a_s^i h^c_{d,i}(\omega,N_1) + (N_1 \leftrightarrow N_2) \,,
\end{align}
where 
\begin{align}
\label{hg}
        h^c_{d,0}(\omega,N_l) &= h^c_{d,00}(\omega) + h^c_{d,01}(\omega) \ln N_l,
        \nonumber\\
         h^c_{d,i}(\omega,N_l) &= \sum_{k=0}^{i} h^c_{d,ik}(\omega)~ \ln^k N_l ,
\end{align}
where $\omega = a_s \beta_0 \ln N_1 N_2$. The SV resummation coefficients, which comprises of $\tilde g_{d,0}^c$ and $g_{d,i}^c$ are greatly discussed in \cite{Banerjee:2017cfc,Banerjee:2017ewt,Ahmed:2020caw} 
and so from here onwards  we focus on the NSV resummation coefficients namely $\overline{g}_{d,i}^c$ and $h^c_{d,i}$. In $\vec N $ space, the use of resummed $a_s$ allows us to organise the series in such a way that $\omega$ is treated as order one at every order in $a_s$. The coefficient $\overline g_{d,1}^c$ is found to be zero. The coefficients $\overline g_{d,i+2}^c$ are controlled by the universal cusp anomalous dimension $A^c$, while $h_{d,i}^c$s  by the NSV coefficients $\varphi_{d,c}^f$ as well as by $C^c, D^c$ from $\mathcal P^c(a_s,\overline z_l)$.  The resummation coefficients $\tilde g^c_{d_0,i},g^c_{d,i}(\omega),\overline {g}^c_{d,i}(\omega)$ and $h^c_{d,i}(\omega)$ encode the entire all order information in a systematic fashion through leading, next-to-leading, $\cdots$, 
SV and NSV logarithms present in the $\Psi_d^c$. For instance, the knowledge of second order  resummation coefficients, $\big\{\tilde g^c_{d_0,0},g^c_{d,1},g^c_{d,2},\bar{g}^c_{d,1}, \bar{g}^c_{d,2} ,h^c_{d,0}, h^c_{d,1}\big\}$, is sufficient to predict the $\ln^{(2i-1)}N_l \over N_l$ of $\Delta_{d,\vec N}^{c,(i)}$ for $i> 2$ to all orders. We present Table [\ref{tab:Table1}] towards the end which demonstrate this feature for  $(\ln^k N_l/N_l)$ terms. In summary, we study the all order logarithmic structure of the NSV terms in $\vec N$ space and the resummation coefficients till 4-loop are provided in the Supplementary Material \cite{footnote:3xxx}.\\
 \textit{Results.}--- We present the third order NSV results for dCFs, $\Delta_{d,c}$, with $c=q,b$, corresponding to DY process and for bottom quark induced Higgs production after expanding them as
$\Delta_{d,c}=~\sum_{i=0}^{\infty} a_s^i \big(\Delta_{d,c}^{\rm SV,(i)} +~\Delta_{d,c}^{\rm NSV,(i)} + \cdots\big)$.  
We have set $\mu_R^2=\mu_F^2=q^2$ and express the results in terms of $SU(N_c)$  Casimirs, namely  $C_F=(N_c^2-1)/2 N_c$ and $C_A=N_c$ and $n_f$, the number of active quark flavours.
\input{DeltaNSV3.tex}
Here, $L_{z_1}=\ln(\overline z_1)$,$\overline \delta=\delta(\overline z_2)$, $\overline {\cal D}_j = \bigg(\frac{\ln^j(\overline z_2)}{(\overline z_2)}\bigg)_+$ and $\zeta_2 = 1.6449\cdots$ and $\zeta_3 = 1.20205\cdots$. Complete third order results for the Higgs production in gluon fusion are already known \cite{Dulat:2018bfe,Lustermans:2019cau}, however we can not confirm our results, which is given in \cite{footnote:3xxx}, with them as they are not publicly available. For the DY, we have found that our third order
prediction is in complete agreement with the \cite{Lustermans:2019cau} for terms of the type ${\cal D}_i(z_l) \ln^j(\overline z_m)$, $i,j\ge 0 , l,m=1,2$. The remaining $\delta(\overline z_l) \ln^j(\overline z_m)$ terms in DY and the complete NSV predictions for Higgs production in bottom quark  annihilation channel at third order are new. Using results up to third order, we can predict three highest NSV logarithms to all order.  Here, the results at fourth order for  
$\ln^j(\overline z_m), j=7,6,5$ are presented.
\input{DeltaNSV4.tex}
This way, we can predict most of the leading NSV terms to all orders in $a_s$. In fact,
the resummation in $\vec N$ space organises SV and NSV threshold logarithms to all orders and the resulting resummation coefficients are controlled by  anomalous dimensions as well as $\varphi_{d,c}^f$ known to a specific order.  The knowledge of these coefficients to specific order in $a_s$ is sufficient to predict the infinite tower of SV and NSV logarithms to a specific accuracy. We summarise our findings in Table[\ref{tab:Table1}]. 
\begin{table}[h!]
\begin{small}
{\renewcommand{\arraystretch}{2}
\begin{tabular}{P{3.2cm}P{.2cm}P{1.4cm}P{1.4cm}P{1.7cm}}
 \hline
 \hline
 \multicolumn{1}{c}{GIVEN} & & \multicolumn{3}{c}{PREDICTIONS}\\
 \cline{1-1}\cline{3-5}
 Resummation Coefficients & &$\Delta^{c,(2)}_{d,\vec N}$ & $\Delta^{c,(3)}_{d,\vec N}$& $\Delta^{c,(i)}_{d,\vec N}$\\
 \hline
 $\tilde g^c_{d_0,0},g^c_{d,1},g^c_{d,2},\bar g^c_{d,1},\bar g^c_{d,2}$,
 $ h^c_{d,0}, h^c_{d,1}$ &  & $ \dfrac{\ln^3 N_l}{N_l}$ &$ \dfrac{\ln^5N_l}{N_l}$ & $\dfrac{\ln^{(2i-1)}N_l}{N_l}$ \\
 $\tilde g^c_{d_0,1},g^c_{d,3},\bar g^c_{d,3},h^c_{d,2}$ &  & &$ \dfrac{\ln^4 N_l}{N_l}$&$ \dfrac{\ln^{(2i-2)}N_l}{N_l}$\\
 $\tilde g^c_{d_0,n-1},g^c_{d,n+1}$, $\bar g^c_{d,n+1},h^c_{d,n}$ &
 & & &$ \dfrac{\ln^{(2i-n)}N_l}{N_l}$  \\
 \hline 
 \hline
\end{tabular}}
\caption{\label{tab:Table1} The all order predictions for NSV logarithms in  $\Delta_{d, \vec N}^{c,(i)}$  for a given set of resummation coefficients }
\end{small}
\end{table}
The results for dCFs and the resummation coefficients are provided in Supplemental Material \cite{footnote:3xxx}.

\textit {Summary}.---Using factorisation property and RG invariance of partonic dCFs, we find that, in addition to the SV terms, the NSV contributions also exponentiate for rapidity  distributions. The perturbative structure of NSV terms for differential distribution with respect to rapidity are then greatly analysed for DY and Higgs productions to all orders. 
Also, the all order structure is  manifested through an integral representation in $z_l$ space, which is used to resum the large logarithms in two dimensional Mellin space in  terms of  $\omega$. This allows one to investigate their numerical impact. Our result expressed in two dimensional $z_l$ space can be used to obtain leading  SV as well as NSV terms to all orders from the lower order results as well as  from inclusive reactions. We present the first results for NSV terms of rapidity distributions till third order for DY \cite{Lustermans:2019cau} and Higgs boson in bottom quark annihilation. From the inclusive results known up to third order in $a_s$, we also predict the leading NSV terms to fourth order for the rapidity distributions of DY  and also for  Higgs productions in both bottom quark annihilation and gluon fusion for the first time. The entire set up advocated in this letter is for the study of diagonal partonic channels 
can also be suitably extended to investigate the all order structure of other potential non-diagonal partonic channels as well.\\
\textit {Acknowledgements} --- We thank J. Michel and F. Tackmann for
third order DY results of rapidity for comparing purposes and 
C. Duhr and B. Mistlberger for providing third order results for the inclusive reactions. We also thank P. Mukhopadhyay for useful discussions.  
\bibliography{xnsv}
\end{document}

%% file: DeltaNSV3.tex
\begin{widetext}
\begin{small}
\begin{align}
\begin{autobreak}
  \Delta_{d,q}^{\rm{NSV},(3)}  = 
         {\color{black}{C_F^3}}   \bigg\{ 
         {\color{black}{L^{5}_{z_1}}}   \big( 
          - 8 \bar{\delta} \big) 
       + {\color{black}{L^{4}_{z_1}}}   \big( 
            44 \bar {\delta}
          - 40 \overline{\cal D}_0  \big) 
       + {\color{black}{L^{3}_{z_1}}}   \Big[ 
           \bar {\delta} \big(
            132 
          + 32 \zeta_2 \big) 
          + 160 \overline{\cal D}_0
          - 160 \overline{\cal D}_1  \Big] 
       + {\color{black}{L^{2}_{z_1}}}
       \Big[- \bar {\delta} \Big(  \frac{1136}{3}  
       + 320 \zeta_3  
       + 96 \zeta_2 \Big) 
          + \overline{\cal D}_0 \big(
            416 
          + 96 \zeta_2 \big)
          + 416 \overline{\cal D}_1
          - 240 \overline{\cal D}_2
           \Big] 
       + {\color{black}{L_{z_1}}}   \Big[ 
            \bar {\delta} \Big(
            848 \zeta_3 
          - \frac{1675}{3} 
          - \frac{88}{3} \zeta_2 
          + \frac{192}{5} \zeta_2^2 \Big)
          - \overline{\cal D}_0 \big(
            640
          + 640 \zeta_3 
          + 192 \zeta_2 \big) 
          + \overline{\cal D}_1 \big(  872 
          + 192 \zeta_2 \big)
                    + 336 \overline{\cal D}_2
          - 160 \overline{\cal D}_3 
        \Big]
       + \Big[ 
           \bar{ \delta} \Big(
            \frac{557}{2} 
          - 384 \zeta_5 
          + 496 \zeta_3  
          + \frac{700}{3} \zeta_2 
          + 128 \zeta_2 \zeta_3 
          - \frac{560}{3} \zeta_2^2 \Big)
          - \overline{\cal D}_0 \Big( 697 
          - 816 \zeta_3  
          - 64 \zeta_2  
          - \frac{192}{5} \zeta_2^2 \Big)
          - \overline{\cal D}_1 \big(
            384 
          + 640 \zeta_3 
          + 288 \zeta_2 \big)
           + \overline{\cal D}_2 \big(
            456 
          + 96 \zeta_2 \big)
          + 80 \overline{\cal D}_3 
          - 40 \overline{\cal D}_4
          \Big]  \bigg\}
       + {\color{black}{C_F^2 n_f}}   \bigg\{ {\color{black}{L^{4}_{z_1}}}   \Big(- \frac{40}{9} \bar {\delta}   \Big) 
       + {\color{black}{L^{3}_{z_1}}}   \Big(\frac{1040}{27} \bar {\delta} 
          - \frac{160}{9} \overline{\cal D}_0
          \Big)
       + {\color{black}{L^{2}_{z_1}}}   \Big[ 
            \bar {\delta} \Big(
            32 \zeta_2
          - \frac{620}{9} \Big)
          + 112 \overline{\cal D}_0 
          - \frac{160}{3} \overline{\cal D}_1
          \Big] 
       + {\color{black}{L_{z_1}}}   \Big[ 
          - \bar {\delta} \Big(
            \frac{9080}{81} 
          + \frac{320}{3} \zeta_3 
          + \frac{32}{3} \zeta_2 \Big) 
          - \overline{\cal D}_0  \Big(   \frac{1040}{9} 
          - 64 \zeta_2 \Big)
          + \frac{640}{3} \overline{\cal D}_1
          - \frac{160}{3} \overline{\cal D}_2  
          \Big] 
       + \Big[ 
           \bar{ \delta} \Big(
            \frac{1999}{27}
          + \frac{2032}{9} \zeta_3 
          - \frac{664}{9} \zeta_2 
          + \frac{256}{15} \zeta_2^2 \Big) 
          - \overline{\cal D}_0 \Big( \frac{1448}{9} 
          + \frac{320}{3} \zeta_3 
          + \frac{32}{9} \zeta_2 \Big)
          + \overline{\cal D}_1 \Big(
            64 \zeta_2 
          - \frac{200}{3} \Big) 
          + 96 \overline{\cal D}_2
          - \frac{160}{9} \overline{\cal D}_3
          \Big] 
        \bigg\}
       + {\color{black}{C_A C_F^2}}    \bigg\{
         {\color{black}{L^{4}_{z_1}}}   \Big( 
            \frac{220}{9} \bar{\delta}
          \Big)  
       + {\color{black}{L^{3}_{z_1}}}   \Big[ 
           \bar {\delta} \Big(
            32 \zeta_2 
          - \frac{5756}{27} \Big)  
          + \frac{880}{9} \overline{\cal D}_0
          \Big] 
       + {\color{black}{L^{2}_{z_1}}}   \Big[  \bar {\delta} \Big(
            \frac{3572}{9} 
          - 168 \zeta_3 
          - \frac{812}{3} \zeta_2 \Big)
          + \overline{\cal D}_0 \big(
            96 \zeta_2      
          - 640 \big) 
          + \frac{880}{3} \overline{\cal D}_1
          \Big] 
       + {\color{black}{L_{z_1}}}   \Big[ 
            \bar {\delta} \Big(
            \frac{70763}{81} 
          + 424 \zeta_3   
          + \frac{20}{3} \zeta_2 
          + \frac{48}{5} \zeta_2^2 \Big)
          + \overline{\cal D}_0 \Big( \frac{6068}{9} 
          - 336 \zeta_3 
          - 512 \zeta_2 \Big)
          + \overline{\cal D}_1 \Big(
            192 \zeta_2  
          - \frac{3784}{3} \Big)
                    + \frac{880}{3} \overline{\cal D}_2     
          \Big] 
       +  \Big[ 
           \bar  {\delta} \Big( 
            \frac{2260}{9} \zeta_2 
          - \frac{56101}{54}
          - 116 \zeta_3 
          + 16 \zeta_2 \zeta_3  
          + 24 \zeta_2^2 \Big) 
          + \overline{\cal D}_0 \Big( \frac{11351}{9} 
          + \frac{728}{3} \zeta_3 
          - \frac{1456}{9} \zeta_2 
          + \frac{48}{5} \zeta_2^2 \Big)  
          + \overline{\cal D}_1 \Big( 
            \frac{1088}{3} 
          - 336 \zeta_3 
          - 448 \zeta_2 \Big)
          + \overline{\cal D}_2 \big(
            96 \zeta_2 
          - 592 \big)
           + \frac{880}{9} \overline{\cal D}_3
          \Big]  
        \bigg\} 
       + {\color{black}{C_A C_F n_f}}   \bigg\{ {\color{black}{L^{3}_{z_1}}}   \Big( 
            \frac{176}{27} \bar{ \delta}
          \Big)
       + {\color{black}{L^{2}_{z_1}}}   \Big[  \bar {\delta} \Big(
            \frac{16}{3} \zeta_2 
          - \frac{1678}{27} \Big)
          + \frac{176}{9} \overline{\cal D}_0 
          \Big] 
       + {\color{black}{L_{z_1}}}   \Big[ 
           \bar {\delta} \Big(
            \frac{14648}{81} 
          - \frac{212}{3} \zeta_2 \Big)   
          + \overline{\cal D}_0 \Big(
            \frac{32}{3} \zeta_2 
          - \frac{3536}{27} \Big) 
          + \frac{352}{9} \overline{\cal D}_1
          \Big]
       + \Big[ \bar  {\delta} \Big(
            \frac{196}{3} \zeta_3 
          - \frac{118984}{729} 
          + \frac{11816}{81} \zeta_2 
          - \frac{208}{15} \zeta_2^2 \Big)
          + \overline{\cal D}_0 \Big( 
            \frac{16952}{81} 
          - \frac{608}{9} \zeta_2 \Big)
          + \overline{\cal D}_1 \Big(
            \frac{32}{3} \zeta_2 
          - \frac{3896}{27} \Big) 
           + \frac{176}{9} \overline{\cal D}_2 
          \Big] 
        \bigg\}
       + {\color{black}{C_F n_f^2}}  \bigg\{
         {\color{black}{L^{3}_{z_1}}}  \Big(
          - \frac{16}{27} \bar {\delta}~
         \Big)
       + {\color{black}{L^{2}_{z_1}}}    \Big( 
            \frac{152}{27} \bar {\delta} 
          - \frac{16}{9} \overline{\cal D}_0
          \Big)
       + {\color{black}{L_{z_1}}}   \Big[ 
           \bar {\delta} \Big(
            \frac{32}{9} \zeta_2  
          - \frac{1264}{81} \Big) 
          + \frac{304}{27} \overline{\cal D}_0
          - \frac{32}{9} \overline{\cal D}_1
          \Big] 
       + \Big[ \bar {\delta} \Big( \frac{10856}{729} + \frac{32}{27} \zeta_3 
          - \frac{304}{27} \zeta_2 \Big)
          + \overline{\cal D}_0 \Big(
            \frac{32}{9} \zeta_2  
          - \frac{1264}{81} \Big)
          + \frac{304}{27} \overline{\cal D}_1
          - \frac{16}{9} \overline{\cal D}_2\Big]  \bigg\}  
       + {\color{black}{C_A^2 C_F}}   \bigg\{
         {\color{black}{L^{3}_{z_1}}}   \Big( 
          - \frac{484}{27} \bar{\delta}
          \Big)
       + {\color{black}{L^{2}_{z_1}}}   \Big[ \bar {\delta} \Big(\frac{4676}{27} 
          - \frac{98}{3} \zeta_2 \Big)   
          - \frac{484}{9} \overline{\cal D}_0 
          \Big] 
       + {\color{black}{L_{z_1}}}   \Big[ 
           \bar {\delta} \Big(
            \frac{2560}{9} \zeta_2 
          - \frac{47386}{81} 
          + 200 \zeta_3 
          - \frac{176}{5} \zeta_2^2 \Big) 
          + \overline{\cal D}_0 \Big(
            \frac{9496}{27} 
          - \frac{176}{3} \zeta_2 \Big)
          - \frac{968}{9} \overline{\cal D}_1  \Big]
       +\Big[   \bar {\delta} \Big(
            \frac{587684}{729} 
          + 192 \zeta_5  
          - \frac{21692}{27} \zeta_3 
          - \frac{40844}{81} \zeta_2 
          + \frac{176}{3} \zeta_2 \zeta_3     
          + \frac{656}{15} \zeta_2^2 \Big) 
          - \overline{\cal D}_0 \Big(
            \frac{49582}{81} 
          - 176 \zeta_3 
          - \frac{856}{3} \zeta_2 
          + \frac{176}{5} \zeta_2^2 \Big)
          + \overline{\cal D}_1 \Big(\frac{11476}{27}
          - \frac{176}{3} \zeta_2 \Big)  - \frac{484}{9} \overline{\cal D}_2 \Big] \bigg\}   +  \big( z_1 \leftrightarrow z_2 \big)\,,
\end{autobreak}    
\nonumber \\
\begin{autobreak}
    \Delta_{d,b}^{\rm NSV,(3)}  =
        \Delta_{d,q}^{NSV,(3)}  + \bigg[
        {\color{black}{C_F^3}} \Big\{ 
         {\color{black}{L^3_{z_1}}}   \big(  
           -96 \bar{\delta} \big)  
         + {\color{black}{L^2_{z_1}}}  \big(  
           288 \bar{\delta}
         - 288 \overline{{\cal D}}_0   
          \big)
         + {\color{black}{L_{z_1}}}  \Big[   
          \bar{\delta}   \big( 471 - 88 \zeta_2 \big) 
          + 480 \overline{{\cal D}}_0  
         - 576 \overline{{\cal D}}_1   \Big] 
       + \Big[ -  \bar{\delta}   \Big(   \frac{447}{2} 
       + 384 \zeta_3 
        148 \zeta_2 \Big)
       + \overline{{\cal D}}_0 \big(  591 - 88 \zeta_2 \big)
       + 288 \overline{{\cal D}}_1 
        - 288 \overline{{\cal D}}_2  
         \Big] 
         \Big\} 
      + {\color{black}{C_F^2}} {\color{black}{n_f}}
       \Big\{   {\color{black}{L^2_{z_1}}} \big(  
        - 16 \bar{\delta}  \big)
       + {\color{black}{L_{z_1}}} \Big[\bar{\delta}  \Big( \frac{1642}{9} - 32 \zeta_2 \Big)
       - 32 \overline{{\cal D}}_0 
      \Big]
         +\Big[ - \bar{\delta}   
         \Big(
          \frac{479}{3} 
         - 48 \zeta_2 \Big)
         + \overline{{\cal D}}_0   \Big( \frac{1642}{9}
         - 32 \zeta_2  \Big)
        - 32 \overline{{\cal D}}_1  \Big]\Big\}
      +{\color{black}{C_A}} {\color{black}{C_F^2}} 
       \Big\{ {\color{black}{L^2_{z_1}}}   88 \bar{\delta} 
        +  {\color{black}{L_{z_1}}} \Big[  \bar{\delta}   \Big( 144 \zeta_3 + 256 \zeta_2 - \frac{9925}{9} \Big)  +    176   \overline{{\cal D}}_0  \Big]
        + \Big[  \bar{\delta}   \Big( \frac{4615}{6} - 408 \zeta_3 - 304 \zeta_2 \Big) - \overline{{\cal D}}_0\Big(\frac{10861}{9} - 144 \zeta_3 - 256 \zeta_2  \Big)  +176 \overline{{\cal D}}_1  \Big] \Big\} + {\color{black}{C_A^2}}  {\color{black}{C_F}} \Big\{ {\color{black}{L_{z_1}}} ~ 8  \bar{\delta} -  \Big[ 16 \bar{\delta} \Big]  \Big\} + \big( z_1 \leftrightarrow z_2 \big) \bigg]\,.
\end{autobreak}
\end{align}
\end{small}
\end{widetext}

%% file: DeltaNSV4.tex
\begin{widetext}
\begin{small}
\begin{align}
\begin{autobreak}
  \Delta_{d,q}^{\rm NSV,(4)}  =
  
     {\color{black}{C_F^4 }}   \Big\{ 
     {\color{black}{L^7_{z_1}}} \Big(- \frac{16}{3} \bar {\delta}  \Big)
     + {\color{black}{L^6_{z_1}}} \Big(  
         \frac{128}{3} \bar {\delta} 
       - \frac{112}{3} \overline{{\cal D}}_0\Big)  
       +   {\color{black}{L^5_{z_1}}} \Big[   \bar {\delta} \big(  132 + 96 \zeta_2  \big)   
      
       +  240 \overline{{\cal D}}_0  - 224 \overline{{\cal D}}_1  \Big]  \Big\}    
       
   + {\color{black}{C_F^3 n_f  }}   \Big\{ {\color{black}{L^6_{z_1}}}\Big(  - \frac{56}{9} \bar{\delta} \Big)

       +   {\color{black}{L^5_{z_1}}} \Big(  \frac{1864}{27} \bar {\delta} - \frac{112}{3} \overline{{\cal D}}_0  \Big) \Big\} 
       
   +  {\color{black}{C_A C_F^3}} \Big\{ {\color{black}{L^6_{z_1}}}  \Big(  \frac{308}{9} \bar {\delta} \Big)

       +  {\color{black}{L^5_{z_1}}}  \Big[  \bar {\delta} \Big(- \frac{10576}{27}  + 48 \zeta_2  \Big) + \Big( \frac{616}{3} \overline{{\cal D}}_0\Big) \Big]\Big\}
       
   +  {\color{black}{C_F^2 n_f^2}} \Big\{ {\color{black}{L^5_{z_1}}}  \Big( - \frac{64}{27} \bar {\delta}\Big) \Big\}

       +   {\color{black}{C_A C_F^2 n_f}} \Big\{
       {\color{black}{L^5_{z_1}}}   \Big(\frac{704}{27} \bar {\delta} \Big)  \Big\}

       +  {\color{black}{C_A^2 C_F^2}} \Big\{  {\color{black}{L^5_{z_1}}}    \Big(  - \frac{1936}{27}  \bar {\delta} \Big) \Big\} + \mathcal{O}\Big( {\color{black}{ L^4_{z_1}}} \Big) +  \big( z_1 \leftrightarrow z_2 \big)\,,
    
\end{autobreak}
\nonumber \\

\begin{autobreak}
  
\Delta_{d,b}^{\rm NSV,(4)} = 

  \Delta_{d,q}^{NSV,(4)} + \Big\{ {\color{black}{C_F^4}} \big[ {\color{black}{L^5_{z_1}}} \big(  - 96 \bar{\delta}~ \big) \big] + \mathcal{O}\Big( {\color{black}{ L^4_{z_1}}} \Big) +  \big( z_1 \leftrightarrow z_2 \big)  \Big\} \,,  
  
\end{autobreak}
\nonumber \\
\begin{autobreak}
  
 \Delta_{d,g}^{\rm NSV,(4)}  =
 
 {\color{black}{C_A^4}}   \Big\{
            {\color{black}{L^{7}_{z_1}}} \Big( 
          - \frac{16}{3}  \bar {\delta} \Big)
          + {\color{black}{L^{6}_{z_1}}} \Big[
            \frac{692}{9} \bar {\delta}
          - \frac{112}{3} \overline{\cal D}_0 \Big] 
          + {\color{black}{L^{5}_{z_1}}} \Big[  \bar {\delta} \Big( 144 \zeta_2 - \frac{12224}{27} \Big) +
            \frac{1336}{3}  \overline{\cal D}_0
          
          - 224 \overline{\cal D}_1 
          \Big]
          \Big\} 
          
  + {\color{black}{C_A^3 n_f}}    \Big\{
          {\color{black}{L^{6}_{z_1}}} \Big(
        -  \frac{56}{9} \bar {\delta} \Big) 
        + {\color{black}{L^{5}_{z_1}}} \Big[ 
          \frac{796}{9}  \bar {\delta}
        - \frac{112}{3}  \overline{\cal D}_0 \Big]
          \Big\}
          
  + {\color{black}{C_A^2 n_f^2}} \Big\{
            {\color{black}{L^{5}_{z_1}}} \Big( 
          - \frac{64}{27}  \bar {\delta} \Big) 
          \Big\} + \mathcal{O}\Big( {\color{black}{ L^4_{z_1}}} \Big)
          
   +  \big( z_1 \leftrightarrow z_2 \big).
\end{autobreak}
\end{align}
\end{small}
\end{widetext}